# Discovery of Ferroelectric Twin Boundaries in a Photoactive Halide Perovskite

Weilun Li[1], Qimu Yuan[2], Michael B. Johnston[2], Joanne Etheridge[1,3,4*]

[1]School of Physics and Astronomy, Monash University, Victoria, 3800, Australia

[2]Department of Physics, University of Oxford, Oxford OX1 3PU, United Kingdom

[3]Monash Centre for Electron Microscopy, Monash University, Victoria, 3800, Australia

[4]Department of Materials Science and Engineering, Monash University, Victoria, 3800, Australia

*e-mail: joanne.etheridge@monash.edu

**Abstract**

Halide perovskites have emerged as promising materials for next-generation photovoltaics, laser sources and X-ray detectors. There is intense debate as to whether some photoactive halide perovskites exhibit ferroelectric behaviour and whether it might be possible to utilise the bulk photovoltaic effect to enhance the performance of halide perovskite solar cells. Here, using low-dose scanning transmission electron microscopy, we discover the existence of ferroelastic twin boundaries in vapor-deposited $CsPbI_3$ thin films, parallel to {110} and {112}. Remarkably, despite photoactive $CsPbI_3$ being centrosymmetric and non-polar, we observe directly that Pb atoms shift at {110} twin boundaries driving a local ferroelectric-like polarisation. These polar twin walls form an intrinsic array of nanoscale functional interfaces, spaced ~30-50 nm apart, embedded within the non-polar perovskite lattice. In contrast, {112} twin boundaries remain non-polar but strongly suppress octahedral tilt and off-centre Cs atom displacements, revealing a different untapped ferroic degree of freedom. These discoveries together uncover previously hidden ferroic functionality in halide perovskite semiconductors, opening opportunities for enhanced conductivity and photovoltaic behaviour through domain wall engineering.

**Introduction**

Metal halide perovskite (MHPs) semiconductors have attracted intense attention in recent years owing to their remarkable optoelectronic properties and rapid deployment in high-performance, low-cost solar cells[1,2], lasers[3,4], light-emitting diodes[5,6], and photodetectors[7].

A defining feature of the perovskite structure is its ability to host a rich variety of ferroic phenomena, such as ferroelectricity, ferroelasticity, flexoelectricity, enabled by its structurally flexible lattices and propensity to break inversion symmetry[8–10].

In the mature field of *oxide* perovskites, ferroic properties underpin a range of physical behaviours and functional responses which can be manipulated by engineering domains and domain walls, offering pathways to enhance material performance[11,12]. In particular, in solar cells, the ferroelectric effect can enable the bulk photovoltaic effect (BPVE), in which the built-in electric field generates an above-bandgap photovoltage that can overcome the theoretical efficiency limit of standard solar cells[13–16].

In the much younger field of *halide* perovskites, understanding of ferroic behaviour is nascent, with relatively few studies investigating the possible existence of ferroic properties. Ferroelastic domains have been observed in certain systems, such as $CsPbBr_3$[17–19] and $MAPbI_3$[20–22], however, there remains intense debate as to whether ferroelectricity exists in these, or indeed any other, halide perovskite semiconductor[22–28]. Establishing ferroelectric behaviour in solar cells is critical because it has profound implications for the BPVE[13–16], conductivity[11,29,30] and charge separation[31,32], all of which can be strongly influenced by internal electric fields associated with charged domains and domain walls.

A key challenge in identifying ferroelastic and ferroelectric domains in halide perovskites has been the unambiguous determination of crystal symmetry and domain structures at the relevant length scales[22,33,34], while avoiding structural degradation. Techniques such as X-ray diffraction (XRD), electron diffraction, and piezo-response force microscopy (PFM) have revealed mesoscale domain patterns in various halide perovskite systems[18,20,21,23,35]. However, these various observations have yielded different interpretations, with proposed origins including polarisation[23,35], spontaneous strain[18,20,21] and compositional inhomogeneity[22]. Moreover, the limited spatial selectivity of these techniques prevents direct insights into local atomic-scale structures, such as domain walls, which are often atomically coherent or confined to just a few atomic layers. This limitation might be addressed by atomic-resolution transmission electron microscopy (TEM) which has been a powerful tool for elucidating ferroic domain structures in oxide perovskites[36,37]; however, the extreme electron beam sensitivity of halide perovskites poses a significant obstacle to achieving quantitative structural information at the atomic level[38,39]. Consequently, direct atomic-scale evidence of ferroelastic domains in halide perovskites remains scarce[17–19]. **Furthermore, there has, as yet, been no *atomic-scale* evidence of ferroelectric domains.**

All-inorganic $CsPbX_3$ (where X = I⁻, Br⁻ or Cl⁻) have attracted increasing interest due to their remarkable optoelectronic properties combined with improved environmental stability compared to hybrid organic-inorganic counterparts[40–42]. It has been proposed that the lone pair on $Pb^{2+}$ cations can lead to ferroelectricity in $CsPbX_3$ through Pb off-centre displacements[43,44] and nanoscale ferroelectricity has been predicted theoretically[45]. However, apart from a report of a polar orthorhombic phase at freezing temperatures in $CsPbBr_3$ (space group: *Pna2₁* below 263K)[46], the majority of research on $CsPbX_3$ has reported an orthorhombic phase (γ-phase) at room temperature with a centrosymmetric, hence non-polar, structure (space group: *Pnma/Pbnm*)[47,48].

In the important case of $CsPbI_3$, which has a bandgap optimally suited as the intrinsic absorber for both single-junction and multi-junction photovoltaic devices, local structural investigations are limited owing to its extreme phase instability. To the best of our knowledge, no twin domains, let alone ferroelectric domains, have been reported[49,50].

Here, we address these questions by leveraging low-dose atomic-resolution TEM to probe directly the atomic structure of vapor co-deposited $CsPbI_3$ thin films. We confirm that bulk $CsPbI_3$ in the orthorhombic photoactive phase is non-polar, however, we discover the existence of two types of twin domains, bounded by {110} and {112} planes, respectively. Focusing on determining the symmetry, strain, and local atomic positions at these domain interfaces, we identify net charge polarisation necessary for ferroelectric phenomena in this photoactive perovskite material. Specifically, and importantly, a quantitative analysis of atomic positions reveals local polar distortions at the {110} twin boundaries which generate thin charge-polarised domains just 2-4 unit cells wide parallel to the boundary and repeating at ~30-50nm intervals within each grain. In contrast, the {112} twin boundaries do not exhibit such net charge polarisation; instead, they are characterised by relaxation of the octahedral tilts and Cs off-centre displacements. Our findings reveal directly, at the atomic scale, the existence of ferroelectric domains in $CsPbI_3$ which can act as nanoscale functional elements. This underscores the potential of domain-wall engineering as a design principle for perovskite semiconductors.

## Results

### Non-polar bulk $CsPbI_3$

We began our study by determining the crystal structure, symmetry and polarity of bulk $CsPbI_3$. (By 'bulk' we mean in regions of single crystal away from grain boundaries, planar and point defects.) Thin films of $CsPbI_3$ were prepared by thermal vapor co-deposition, a solvent-free, scalable technique widely used for fabricating high-quality MHP devices[51,52].

Using annular dark field scanning transmission electron microscopy (STEM-ADF), we imaged the atomic structure and measured the positions of atomic columns (**Fig. 1(C-J)**), keeping electron dose as low as possible (see Methods). From atomic-resolution STEM-ADF images along multiple zone axes, we confirm that $CsPbI_3$ adopts an orthorhombic perovskite structure, commonly referred to as γ-$CsPbI_3$. We quantitatively analysed Pb and Cs atom positions and found no detectable Pb off-centre displacements anywhere in the single-crystal regions away from domain walls, confirming there is no spontaneous polarisation in the bulk. Consistent with the γ-$CsPbI_3$ phase, Cs shows anti-polar off-centre displacements, that is, Cs atoms on adjacent A-sites shift in opposite directions along the [001] direction (**Fig. S3**). These antipolar displacements, which arise from octahedral tilts rather than spontaneous polarisations, stabilises the octahedral-tilted perovskite structure[53].

Taken together, our quantitative analyses are consistent with bulk orthorhombic $CsPbI_3$ exhibiting a non-polar, centrosymmetric structure. No evidence of Pb off-centring was detected, consistent with the centrosymmetry of the Pb-I octahedra in the *Pbnm* phase and previous reports.

This phase exhibits in-phase octahedral tilts along the long axis (c-axis) and out-of-phase tilts along the other two orthogonal axes, resulting in a distorted Pb-I octahedral network (**Fig. 1A**). These octahedral tilts lead to distinct lattice parameters (*a, b, c*) along the three orthogonal axes and characteristic non-equal axes compared to the high-symmetry phases, approximately satisfying: $a_{ortho} \approx b_{ortho} \approx \sqrt{2}a_c$ and $c_{ortho} \approx 2a_c$, where "ortho" and "c" refer to orthorhombic and cubic unit cells, respectively (**Fig. 1B**).

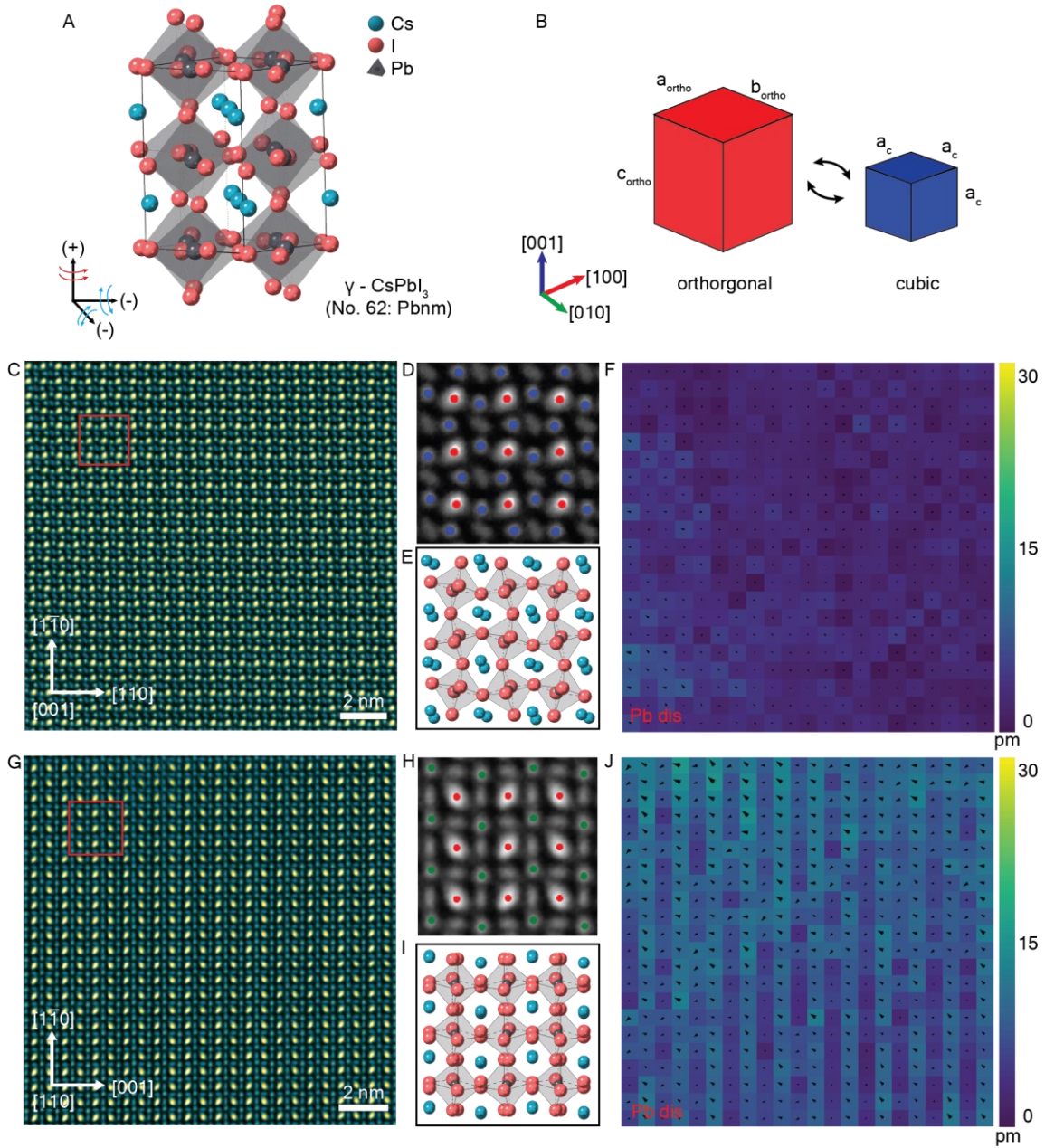

**Fig. 1. Atomic position measurement in bulk γ-CsPI₃.** **(A)** Atomic model of orthorhombic γ-CsPbI₃ (*Pbnm* space group) with octahedral rotations described by the Glazer notation a⁻b⁻c⁺, based on database ICSD-21955[54]. **(B)** Schematic illustrating the relationship between the orthorhombic ("ortho") unit cell and the cubic ("c") unit cell. **(C)** STEM-ADF image of a CsPbI₃ single crystal viewed along the [001] zone axis. **(D)** Enlarged view of the region marked in **(C)**. **(E)** Atomic structural model of orthorhombic phase CsPbI₃ viewed along the [001] zone axis. **(F)** Pb off-centre displacement map derived from atomic-resolution STEM-ADF image in **(C)**, showing the degree of off-centre displacement, if any, of Pb/I columns (red in **(D)**) relative to the geometric centre of the four surrounding I columns (blue in **(D)**). **(G)** STEM-ADF image of a CsPbI₃ grain viewed along the [110] zone axis. **(H)** Enlarged view of the region marked in **(G)**. **(I)** Atomic structural model of orthorhombic phase CsPbI₃ viewed along the [110] zone axis. **(J)** Pb off-centre displacement map derived from atomic-resolution STEM-ADF image in **(G)**, showing the degree of off-centre displacement, if any, of Pb/I columns (red in **(H)**) relative to the centre of the four surrounding Cs columns (green in **(H)**). Vectors in **(F)** and **(J)** indicate the direction and magnitude of Pb atom displacement from centre.

## {110} Ferroelastic Twin Domains

We next extended our structural investigation to twin domains and twin boundaries (i.e. domain walls). To the best of our knowledge, direct observations of twin boundaries have not been reported previously in $CsPbI_3$. A TEM overview image (**Fig. 2A**) reveals a homogeneous grain structure with grain sizes ranging from ~100-200 nm. Notably, some grains exhibit a distinctive "stripe contrast", dividing individual grains into two or more subdomains (additional overview images are shown in **Fig. S4**).

**Fig. 2B** shows a grain viewed along the [001] zone axis, where intra-grain domains are clearly visible, with typical widths of ~30-50 nm. The corresponding Fourier Transform (FT) of the domain region (**Fig. 2C**) exhibits characteristic "double spots" or "twinned spots" (highlighted in **Fig. 2D**), which is a well-known signature of crystallographic twinning in perovskite materials[21,55].

Across the twin boundary, the crystal lattice is mirrored, corresponding to an interchange of the a- and b- axes relative to the (110) twin plane. Since the b-axis is slightly longer than the a-axis, this interchange produces two distinct reciprocal lattice spots, one from each of the two twinned domains: the spots coincide along the twin axis but are spatially offset elsewhere, as is clearly resolved in the FT pattern. The axes interchange results in a lattice rotation across the twin boundary at about 3° measured from the angle between the spots in **Fig. 2D** and diffraction patterns in **Fig. S8H**. This angle is consistent with the lattice spacing ratio of b/a≈1.03. This can be analysed with geometric phase analysis (GPA), which highlights the crystal rotation induced by twinning, as shown in **Fig. 2E**.

The crystallography of the {110} twin boundary in $CsPbI_3$ is illustrated schematically in **Fig. 2F**, along with the geometry of the corresponding diffraction patterns in **Fig. 2G**. These twin boundaries delineate a switch between crystal orientation (described as A-type domains and B-type domains in **Fig. 1**) and are indicative of ferroelastic behaviour in perovskite materials[56,57].

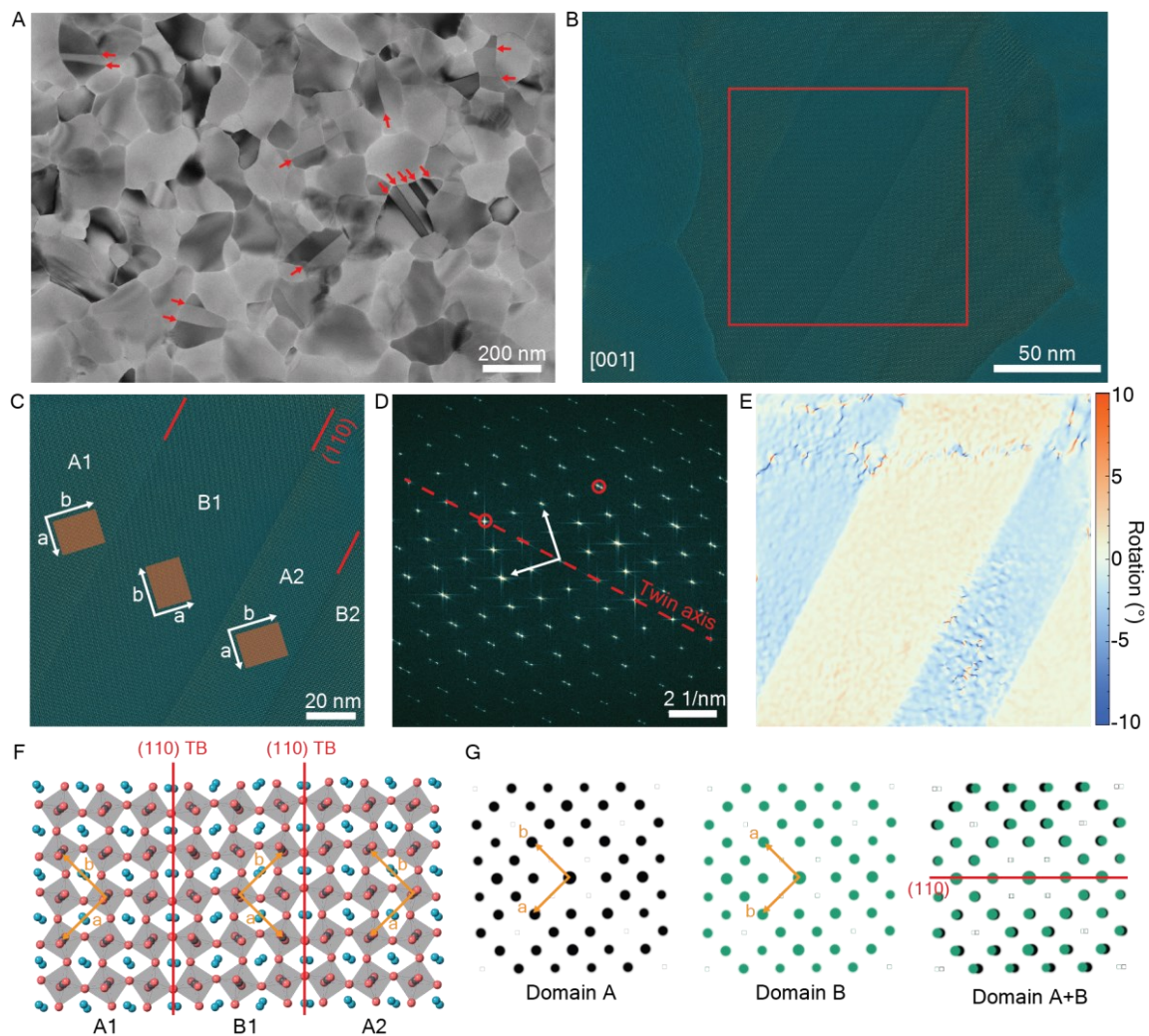

**Fig. 2. {110} twin domains in CsPbI$_3$.** **(A)** TEM overview image of CsPbI$_3$ thin films. **(B)** HR-TEM image of a representative grain viewed along the [001] zone axis (red arrows point to twin domains). **(C)** Enlarged view of the domain region highlighted in **(B)**. **(D)** Fourier Transform of **(C)**. **(E)** Rotation map of the region in **(C)**, derived from GPA of the circled spots **(D)**. **(F)** Schematic of the (110) twin boundary crystallography, viewed along the [001] zone axis. **(G)** Simulated kinematic diffraction patterns of the A and B domains in **(F)**, with the overlaid pattern (A+B) at right. Squares indicate reflections kinematically forbidden by the crystal symmetry. Domains A and B correspond to regions with distinct crystal orientation. Image in (C) was generated by summing 150 frames of ultra-low-dose HR-TEM frames, each acquired at 2 e/Å$^2$, to preserve structural integrity and avoid beam damage (see **Fig. S5**).

## Localised Polarisation at {110} Twin Boundaries

We have shown that orthorhombic CsPbI$_3$ adopts a centrosymmetric structure consistent with the reported *Pbnm* space group. However, we find that this inversion symmetry can be locally broken at and near the twin boundaries, introducing structural degrees of freedom that are absent in the bulk crystal, which we describe here.

A representative region containing a {110} twin boundary is shown in **Fig. 3A**. The interchange of crystallographic axes across the boundary is evident from local measurements of Pb-Pb interatomic distances (corresponding to the *a* and *b* lattice parameter), as quantified in **Fig. 3B**. These structural changes confirm the presence of a (110) twin boundary and are fully consistent with the crystallographic model described earlier in **Fig. 1**. Importantly, this analysis also allows us to pinpoint the precise location of the twin boundary, which is challenging to discern directly from the low-dose TEM image due to subtle intensity variations.

To measure the local atomic positions at the twin boundary, we analysed the position of Pb/I columns relative to the geometric centre of their surrounding I columns (**Fig. 3C**) as described in Methods. Strikingly, we observed significant off-centre displacements of the Pb/I atomic columns localised at the twin boundary, with the displacement vector predominantly parallel to [110], along the boundary plane. These displacements are confined to a narrow region approximately four atomic layers wide (2 layers each side of the boundary), indicating that the boundary itself hosts a distinct structural configuration which is polar. No evidence of polarisation was found within the domain interiors, confirming that they retain the non-polar character of the bulk material.

To further quantify this behaviour, we resolved the Pb displacements into components parallel and perpendicular to the twin boundary and plotted their magnitudes in **Fig 3D**. While minor background variations arising from measurement noise, local crystal bending, and slight off-axis tilt are observed (as demonstrated in **Fig. S6** and **S7**), a pronounced increase in displacement is evident at the twin boundary, relative to the interior. The peak magnitude of the displacement reaches ~ 15 pm at the twin boundary, clearly distinguishing it from the surrounding bulk domains, where displacements remain negligible.

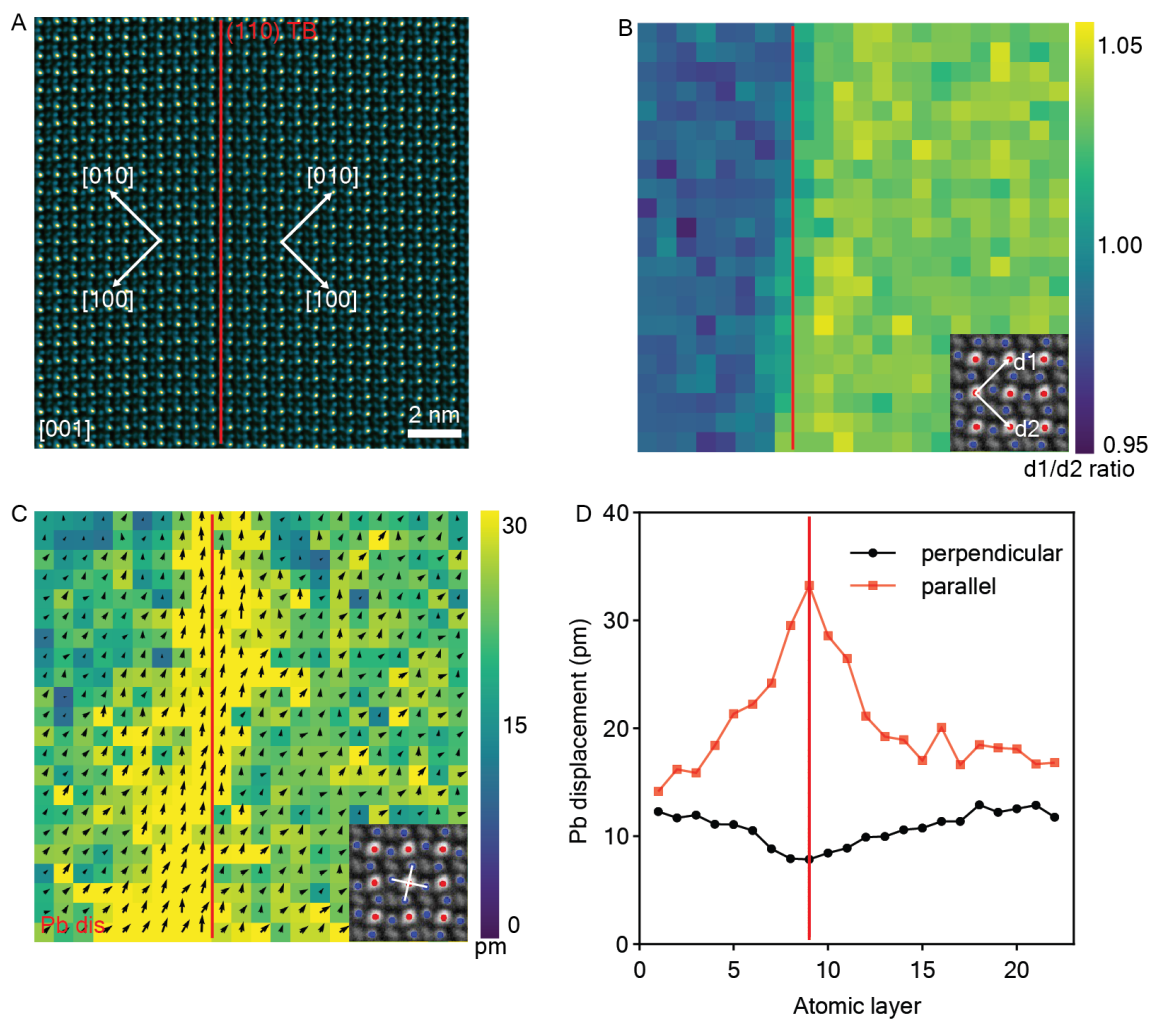

**Fig. 3. Local polarisation at the {110} twin boundary. (A)** Atomic-resolution STEM-ADF image of a [001]-oriented grain containing a (110) twin boundary. **(B)** Ratio of Pb-Pb distances ($d1/d2$) across the twin boundary, as illustrated in the inset. **(C)** Map of Pb/I column off-centre displacements relative to the geometric centre of the four surrounding I columns, as illustrated in the insert. Vectors represent the direction and magnitude of Pb off-centre displacements. **(D)** Components of the Pb displacement vector resolved parallel and perpendicular to the twin boundary, obtained by averaging the displacement map in **(C)**. (A slight mistilt of the crystal away from the zone axis leads to an off-set error of all displacement measurements of ~15 pm in the parallel direction and ~10 pm in the perpendicular direction). Red vertical lines indicate the position of the twin boundary.

**Non-polar {112} Twin Boundaries**

We also identified a second type of twin boundary in CsPbI$_3$ films, orientated parallel to the {112} plane. These {112} twin boundaries are observed only occasionally, likely due to the strong [001]-preferred grain orientation in vapor-deposited CsPbI$_3$ films. In other words, they may occur more frequently than observed but we cannot see them due to the preferred grain orientation.

The {112} twins involve an interchange of the [110] and [001] crystallographic directions across the boundary (as opposed to the [100] and [010] directions for the {110} twin boundaries). **Fig. 4A** shows a [110]-orientated grain investigated using low dose four-dimensional STEM (4D-STEM) (see Methods). The grain is divided into two sub-domains by a twin boundary parallel to the (112) plane. Notably, the diffraction patterns from either side of the boundary (**Fig. 4B** and **4C**) exhibit a clear mirror symmetry in the arrangement of diffraction spots, reflecting the interchange of crystallographic axes across the twin boundary. A comparison of 4D-STEM results for the two types of twins is provided in **Fig. S8**.

An atomic-resolution STEM-ADF image of a typical {112} twin boundary is shown in **Fig. 4D**. Interestingly, no measurable displacement of the Pb/I columns relative to the centres of the surrounding octahedra is observed (**Fig. 4E**), indicating the absence of localised polarisation at {112}-type twin boundaries, in contrast to the polar distortion observed at the {110} twin boundary. Analysis of Cs positions reveals pronounced anti-polar displacements in the interior of the domain, as expected for the non-polar *Pbnm* space group (**Fig. 4F**). Importantly, however, we observe that these Cs displacements reduce significantly in magnitude within the transition region at the boundary.

This suppression of Cs atomic displacements at the {112} twin boundary occurs together with the suppression of octahedral tilts at the boundary, as evidenced by measurements of the "ellipticity" of the intensity maxima associated with the Pb/I columns (**Fig. 4G**). Together, these results suggest that, while the {112} twin boundaries preserve the centrosymmetry of the Pb-I framework, they reduce the antipolar displacements of Cs and the octahedral tilt to accommodate the structural reflection between the adjacent twin domains.

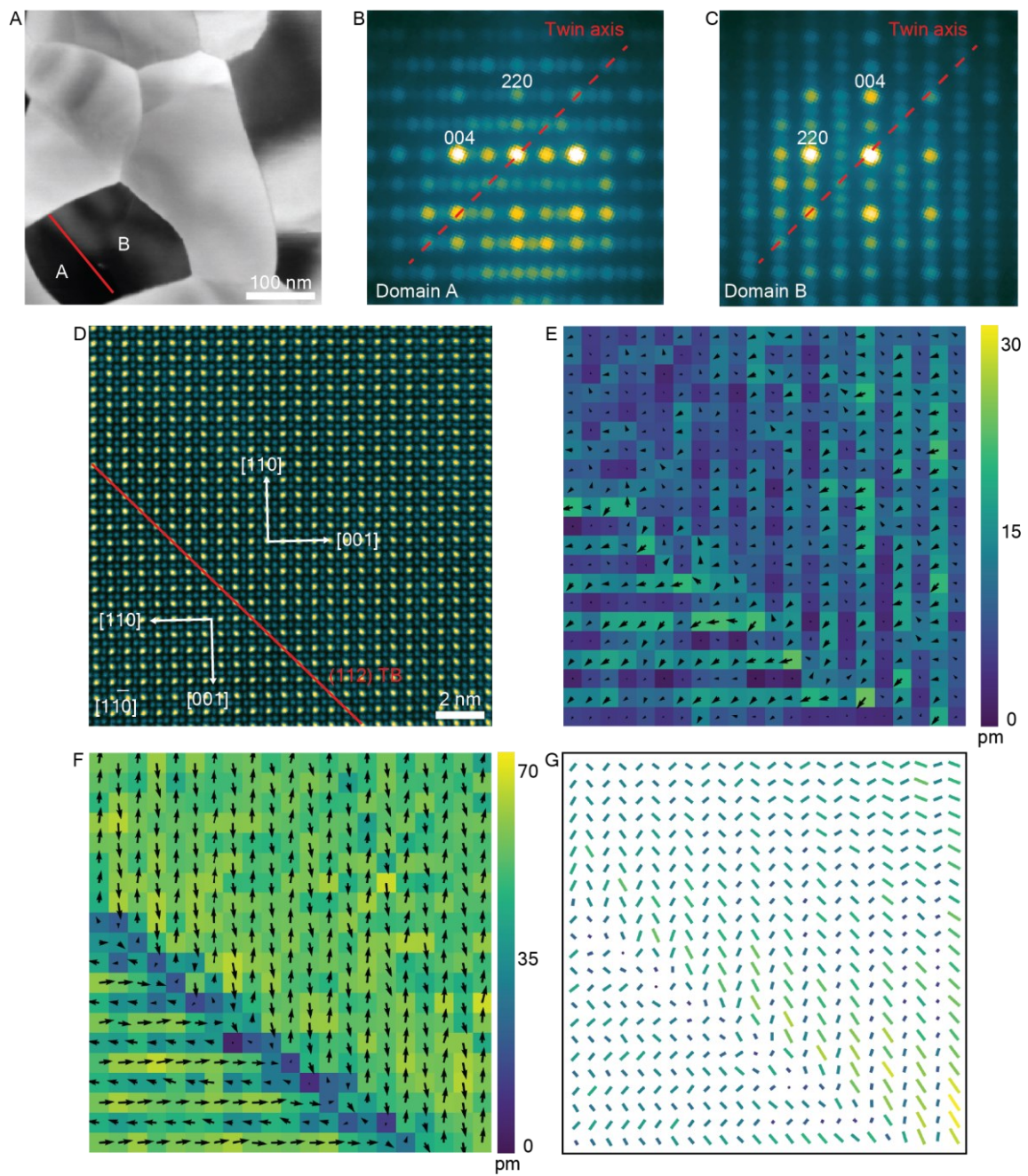

**Fig. 4. Non-polar {112} twin boundary. (A)** Image of a [110]-oriented grain generated by integrating all pixel signals within each diffraction pattern of the acquired 4D-STEM dataset. **(B)** Averaged

diffraction pattern from domain A. **(C)** Averaged diffraction from domain B. **(D)** STEM-ADF image of a representative {112} twin boundary. **(E)** Map of Pb/I column off-centre displacements relative to four surrounding Cs columns, as illustrated in the insert. Vectors indicate the direction and magnitude of Pb off-centre displacements. **(F)** Map of Cs column off-centre displacements relative to four surrounding Pb/I columns, as illustrated in the insert. Vectors indicate the direction and magnitude of the Cs off-centre displacement. **(G)** Ellipticity vector map measured from Pb/I columns in **(D)**, a measure of octahedral tilt. The line direction indicates the orientation of the ellipse major axis. Line length and colour indicate the magnitude of ellipticity (major axis/minor axis).

## Discussion

In this study, we discovered and characterised two distinct types of ferroelastic twin boundaries in orthorhombic $CsPbI_3$, each involving an interchange between two orthogonal crystallographic axes: [100]/[010] for the {110}-type twins and [110]/[001] for the {112}-type twins. As $CsPbI_3$ cools from high temperature, it undergoes a sequence of symmetry-lowering phase transitions from cubic to tetragonal and finally to an orthorhombic phase at room temperature, accompanied by octahedral tilts and spontaneous strain[48,49]. In other perovskite compounds, such phase transitions have been shown to promote the formation of ferroelastic twin domains as a mechanism to minimise elastic energy and accommodate these structural changes[55,57].

Our quantitative atomic-resolution STEM-ADF measurements reveal no evidence of spontaneous polarisation in single crystal $CsPbI_3$, *within the interior* of either type of twin domain, indicating the absence of ferroelectricity in the domain interiors in $CsPbI_3$, consistent with its reported space group Remarkably, however, we find that these ferroelastic twin boundaries can locally break inversion symmetry and host polar distortions, despite the centrosymmetric, non-polar nature of the surrounding interior matrix. This phenomenon is clearly observed at {110}-type twin boundaries (**Fig. 5(A, C)**). While polar ferroelastic domain walls have previously been observed in oxide perovskites such as $CaTiO_3$[58], $SrTiO_3$[59], $GdFeO_3$[60], this work provides the first direct experimental evidence of such polar domain walls in a photoactive halide perovskite, demonstrating they are not exclusive to oxide perovskites. This opens some exciting possibilities in analogy with the oxide perovskites where domain wall functionalities[11,12] have already enabled novel device concepts including diodes[61], memories[62] and tunnel junctions[63]. By revealing intrinsic polar nano-interfaces embedded within the semiconductor lattice, our results position halide perovskites as a new platform for exploring nanoscale multi-ferroic phenomena.

Importantly, this boundary-localised ferroelectricity may enhance conductivity[64] and/or the bulk photovoltaic effect[13,65], as has also been observed in some oxide perovskites[29,30,64,66]. Moreover, recent theoretical studies on perovskite oxides predict that the local polarisation at ferroelastic domain walls is switchable under external fields, opening opportunities for domain-wall engineering in devices[16,67,68].

Our measurements show that the B-site cations ($Pb^{2+}$) are displaced by up to ~ 15 pm at the {110} twin boundaries. This displacement is more than twice that observed for B-site displacement in $CaTiO_3$ (~ 6 pm)[58] but smaller than the A-site polar displacement in $GdFeO_3$ (~ 30 pm)[60]. The polarisation vectors are consistently oriented parallel to the twin boundary; a characteristic shared with the polar domain walls in the oxide perovskite systems. The spatial extent of the polar region, confined to approximately four atomic layers, also aligns well with theoretical predictions for $CaTiO_3$[69,70].

Different behaviour applies to the {112}-type domain boundaries. In orthorhombic bulk $CsPbI_3$, the A-site $Cs^+$ cations exhibit alternating off-centre displacements coupled to octahedral tilts, resulting in no net dipole in the bulk crystal. At the {112}-type twin boundaries, this antipolar alternating pattern is disrupted due to the change in octahedral tilt sense across the boundary, which drives the Cs atoms back to a central position (**Fig. 5B**). As a result, no net polarisation arises from A-site displacements, either within the domain or at the {112}-type domain boundaries. Prior studies have shown that such antipolar A-site displacements in *Pbnm* perovskites inhibits ferroelectricity. Furthermore, in order to introduce polarisation[53] it is necessary to suppress the Cs displacement, for example, by applying an external electric field. Our findings suggest that {112} twin boundaries might serve as initial sites for ferroelectric switching by locally reducing the energy barrier for polarisation, potentially lowering the required field[68,71].

The distinction in polar behaviour between the {110} and {112} twin boundaries originate from their differing atomic structures (**Fig. 5**). The {110} boundaries lie parallel to the octahedral framework (along <001>$_c$), intersecting the iodine columns at the octahedral corners. In contrast, the {112} boundaries intersect the octahedral network at ~ 45°, crossing through the centre of the octahedra (along <011>$_c$). This structural difference imposes distinct constraints on the strain distribution and structural distortions at the twin boundary, leading to different polarisation behaviour. Notably, theoretical calculations have predicted polarisation at {112} twin boundaries in tetragonal CsPbI$_3$ and SrTiO$_3$ (space group *P4/mbm*), where only a single octahedral tilt mode is present[16,72]. These results underscore the critical role of octahedral tilts and A-site antipolar motion in governing the emergence or suppression of twin boundary localised polarisation.

We note that such ferroelastic twin boundaries are not expected in all the photoactive halide perovskites, as their formation depends on symmetry and tilt distortions. For instance, in tetragonal phases (where $a_t=b_t$, "t" for tetragonal), only {112} ferroelastic twin boundaries have been reported, such as those in MAPbI$_3$[21]. While in cubic FAPbI$_3$ (where $a_c=b_c=c_c$, "c" for cubic), only non-ferroelastic, face-sharing twin boundaries have been observed[73,74].

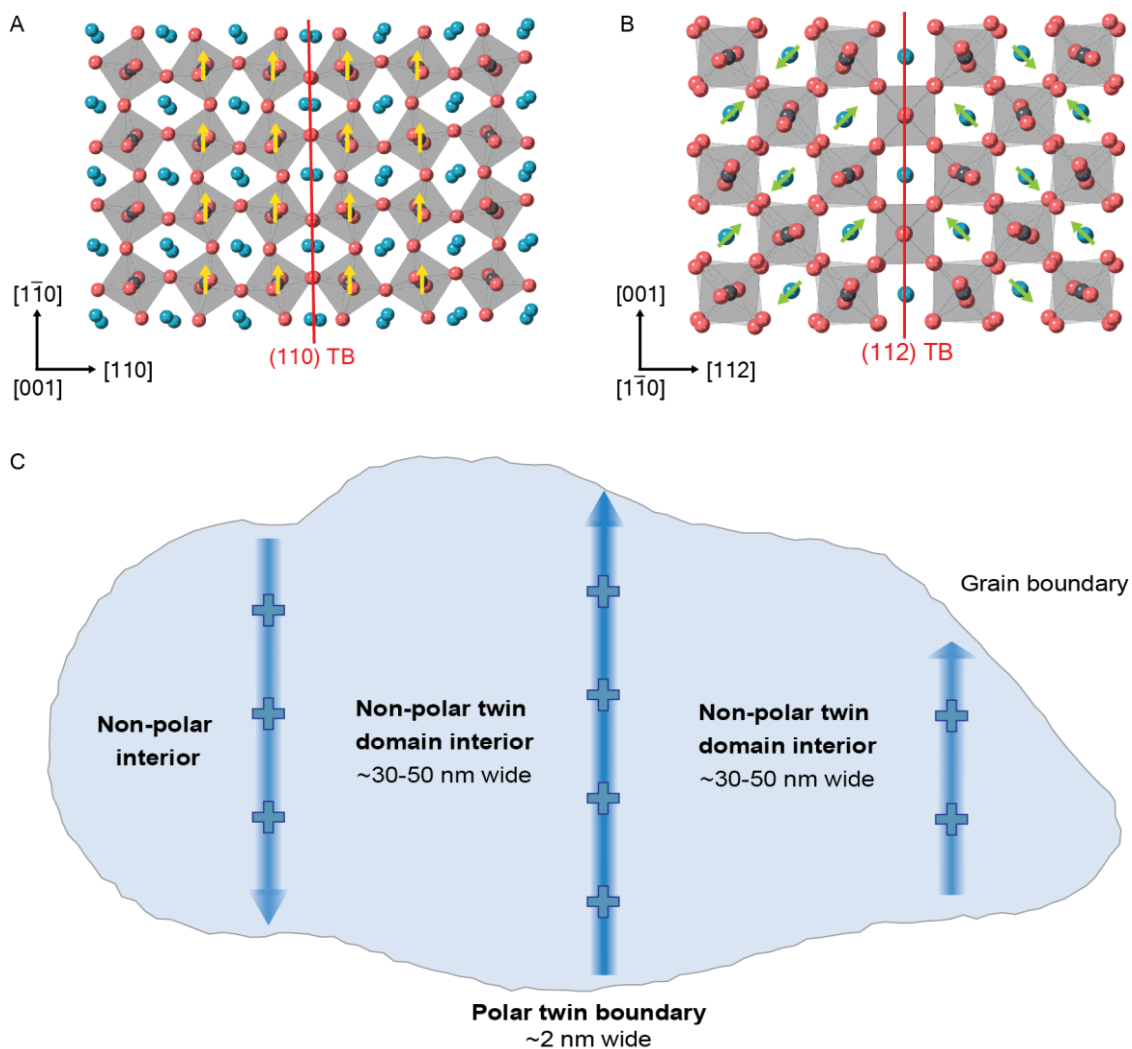

**Fig. 5. Schematics of twin boundaries in orthorhombic CsPbI$_3$.** (**A**) Local polar displacement of Pb at the {110} twin boundary. (**B**) Local suppression of octahedral tilts and Cs antipolar displacement at the {112} twin boundary. (**C**) Schematic of charged {110} twin boundaries embedded within the non-polar bulk lattice.

In summary, we employed low-dose atomic-resolution TEM to reveal and characterise twin domains and domain walls in orthorhombic CsPbI$_3$ thin films, structures not previously observed in this materials class. Importantly, we identify nanoscale polarisation confined to {110} twin boundaries embedded within a non-polar bulk matrix, establishing twin-boundary-induced ferroelectricity as a fundamental ferroic behaviour in halide perovskite semiconductors. Such charged domain walls are known in perovskite oxides to enhance local conductivity and photovoltaic responses, enabling domain-wall-based nanoelectronics[29,30,64,66]. By revealing analogous functionality in the halide perovskite of $CsPbI_3$, a key semiconductor for solar cells and light-emitting devices, our work translates oxide ferroic insight into the realm of halide perovskites. Our findings not only reveal the existence of ferroelectric phenomena in halide perovskites but also establish twin boundaries as a promising platform for domain-wall engineering to improve photophysical properties and in the design of novel perovskite optoelectronic devices.

## Data availability

Source data are provided with this paper in the main text or its Supplementary Information files. Extra data is available from the corresponding authors on reasonable request.

## Acknowledgements

Financial support from the Australian Research Council (ARC) is appreciated. J.E. acknowledges ARC Laureate Fellowship FL220100202 and J.E. and M.B.J. acknowledge ARC Discovery Project DP200103070. The authors acknowledge use of facilities and support of staff within the Monash Centre for Electron Microscopy, a node of Microscopy Australia. The Thermo Fisher Scientific Spectra φ TEM was funded by ARC LE170100118 and the FEI Titan³ 80-300 FEG-TEM was funded by ARC

LE0454166. M.B.J. acknowledge financial support from the Engineering and Physical Sciences Research Council (EPSRC).

## Author contributions

W.L. and J.E. conceived and designed the project. W.L. carried out TEM experiments and analysed data under the supervision of J.E.. Q.Y. optimised and prepared specimens under the supervision of M.B.J.. W.L. and J.E. prepared the manuscript. All authors contributed to the discussion of the results and revision of the manuscript.

## Competing interests

The authors declare no competing interests.

## Additional information

Supplementary information is available.